**On Subthreshold Ionization of Helium Droplets, Ejection of He$^+$, and the Role of Anions**

Michael Renzler,[a] Matthias Daxner,[a] Nikolaus Weinberger,[a] Stephan Denifl,[a] Paul Scheier[*a] and Olof Echt[*ab]

The mechanism of ionization of helium droplets has been investigated in numerous reports but one observation has not found a satisfactory explanation: How are He$^+$ ions formed and ejected from undoped droplets at electron energies below the ionization threshold of the free atom? Does this path exist at all? A measurement of the ion yields of He$^+$ and He$_2^+$ as a function of electron energy, electron emission current, and droplet size reveals that metastable He$^{*-}$ anions play a crucial role in the formation of free He$^+$ at subthreshold energies. The proposed model is testable.

a   Institut für Ionenphysik und Angewandte Physik
    Universität Innsbruck
    Technikerstraße 25
    A-6020 Innsbruck, Austria
    Fax: (+43) 512 507 2932
    E-mail: paul.scheier@uibk.ac.at
b   Department of Physics
    University of New Hampshire
    Durham, NH 03824, USA
    E-mail: olof.echt@unh.edu

Research into helium nanodroplets, originally a scientific niche driven by curiosity about the minimum droplet size that supports superfluidity,[1] has matured to a point where $^4$He droplets provide a novel method to synthesize and characterize unusual molecules, large aggregates in unusual morphologies, metallic foam, or nanowires from a wide range of materials.[2-7] Still, not only do the droplets provide new ways for synthesis but the products also provide new insight into properties of helium droplets. For example, the shape of silver aggregates grown in very large droplets reflects the presence of quantized vortices in superfluid droplets.[7,8]

A topic that has been of interest ever since large helium droplets were efficiently produced in supersonic jets[9] is the mechanism by which droplets become charged by ionizing radiation. How do small He$_n^+$ ions containing as few as two atoms emerge from a very large neutral, undoped droplet?[10,11] How do monomer or dimer ions form when the energy of the ionizing radiation is below their thermodynamic threshold?[12-14] How do large He$_n^-$ cluster anions form upon electron impact?[15-18] What role do metastable electronically excited species play?[12,19] What is the local structure near a positive or negative charge in undoped helium droplets, and how does it compare to that in bulk helium or helium films?[20-23]

Our present work addresses the formation and subsequent ejection of bare He$^+$ from undoped droplets. For ionizing radiation exceeding the ionization threshold of atomic He (24.59 eV) small He$_n^+$ cluster ions ($n > 1$) are thought to result from a two-step mechanism.[12,22,24-28] The process commences with the formation of He$^+$ in the droplet. Direct formation of He$_n^+$ cluster ions ($n > 1$) is disfavored by very small Franck-Condon factors. The hole will hop, on the time scale of femtoseconds, by resonant charge exchange with adjacent helium atoms. After about 10 hops the charge will localize by forming a vibrationally excited He$_2^+$. Its excess energy will be large given the large (about 2.4 eV) dissociation energy of He$_2^+$.[29] This energy would be sufficient to boil off thousands of helium atoms (the bulk cohesive energy of helium is 0.62 meV) but a thermal process appears unlikely; evaporation of even thousands of helium atoms from a primary droplet containing »10$^4$ helium atoms would still result in a helium cluster ion whose size lies outside the range of most mass spectrometers. A more realistic scenario is partial transfer of the vibrational energy of He$_2^+$ to the immediate, strongly bound solvation shell; this energy combined with electrostriction leads to the explosive ejection of a small He$_n^+$ cluster ion with a broad distribution of $n$ that is more or less independent of the neutral droplet size.[10,22,24,27]

The process described above cannot lead to the ejection of $He^+$ from a droplet. Mass spectra of undoped droplets recorded at electron or photon energies exceeding the ionization threshold of atomic helium often show a $He^+$ signal that is as strong or even stronger than that of the most abundant cluster ion, $He_2^+$,[30] but the $He^+$ signal is usually attributed[12,14,32,33] to direct ionization of background gas or helium atoms that are evaporated[34,35] from droplets.

However, $He^+$ has also been observed at subthreshold energies[12,13] where the above explanation fails on energetic grounds. Multiple excitations have been invoked but details have remained murky. If the process involves Penning ionization of $He^*$ by another $He^*$,[14] how does that lead to ejection of $He^+$? Or does $He^+$ result from post-ionization of $He^*$ or $He_2^*$ ejected from a singly excited droplet?[12]

Our experiments reveal a striking similarity in the ion yields of free $He^+$ and $He^-$ ions at subthreshold energies. $He^-$ was recently observed as a product of electron attachment to helium droplets; its resonant formation at 22.0 eV was assigned to formation of $He^* + e^-$ in the droplet, subsequent self-trapping of the electron in a bubble state, and reaction of the bubble with $He^*$ to form $He^{*-}$ in a metastable quartet state.[36] Our present data suggest that $He^+$ is formed in multiply excited droplets by Penning ionization of a highly mobile $He^{*-}$,

$$He^{*-} + He^* \rightarrow He^+ + He + 2e^- \qquad (1)$$

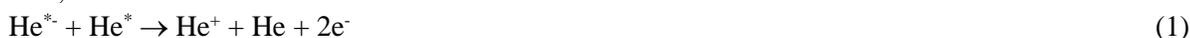

The reaction occurs in the surface region where the heliophobic $He^*$ resides,[37] leading to subsequent ejection of $He^+$. To the best of our knowledge this reaction has not been considered before.[38] The model is supported by the observation that free $He^+$ does not efficiently form below 42 eV if the droplets are too small to support an electron bubble. The validity of the proposed model may be tested by photoionization which cannot produce the required intermediate, $He^* + e^-$.

Neutral helium nanodroplets were produced by expanding pre-cooled helium (purity 99.9999 %) into vacuum. The ensuing droplets are superfluid with a temperature of 0.37 K;[2] we estimate their size[28,40] based on the nozzle diameter (5 µm), stagnation pressure (2.1 MPa) and temperature $T_o$ which was varied from 8 to 11 K. The droplets were exposed to an electron beam for a duration of 3.8 µs, as estimated from the droplet velocity (260 m/s) and electron beam cross section (0.3 × 1.0 mm$^2$). The ensuing ions were accelerated and focused into a commercial reflectron-type time-of-flight mass spectrometer with a resolution $\Delta m/m = 1/2000$. The base pressure in the spectrometer was 10$^{-5}$ Pa; the flight times were 3.3 µs for $He^\pm$ ions. Additional experimental details have been described elsewhere.[36,41]

We have measured the ion yield of $He^+$ and small $He_n^+$ cluster ions as the function of three experimental parameters: energy of the ionizing electrons, electron emission current, and neutral droplet size. Electron current $I_e$ and droplet size $N_{av}$ were varied by more than two orders of magnitude. Fig. 1 displays the ion yields of $He^+$, $He_2^+$ and $He_9^+$ cations *versus* electron energy. Data in panels a and b were recorded for large helium droplets ($T_o$ = 8.5 K, $N_{av} \cong 3\times10^6$ atoms per droplet) and either low or high electron currents (0.4 µA and 60 µA, respectively). Data in panel c cover a wider energy range; they were recorded for small droplets at high current ($T_o$ = 10 K, $N_{av} \cong 2\times10^4$, 100 µA). In the upper two panels the yield of $He_9^+$, which is representative of other small $He_n^+$ cluster ions, is enhanced by a factor 100.

At low electron current (Fig. 1a) the yield of cations rises approximately linearly above an onset near the ionization energy of the helium atom, 24.587 eV.[29] A slight apparent redshift of the onset results from the finite electron energy resolution of 1 eV. It may appear surprising that the apparent threshold of the $He_2^+$ yield agrees with that of $He^+$ in spite of the greatly reduced adiabatic ionization energy of $He_2$ (22.2 eV [29]) but formation of $He_2^+$ involves an intermediate $He^+$; the very large difference between the He-He distance in $He_2^+$ ($r_e = 1.08$ Å) and the condensed phase ($r_e \cong 3.6$ Å) renders the formation of $He_2^+$ by direct ionization highly unlikely.[29,42]

The energy dependence of $He^+$ and $He_2^+$ changes dramatically when the electron current is increased by two orders of magnitude (Fig. 1b). Now the $He^+$ signal exhibits an onset near 21 eV, a maximum at 22.0 eV (the estimated accuracy of the energy calibration is 0.2 eV), a local minimum around 24 eV, and a linear increase beyond 25 eV. $He_2^+$ reaches a plateau at the same energy (22.0 eV) where $He^+$ reaches a local maximum.

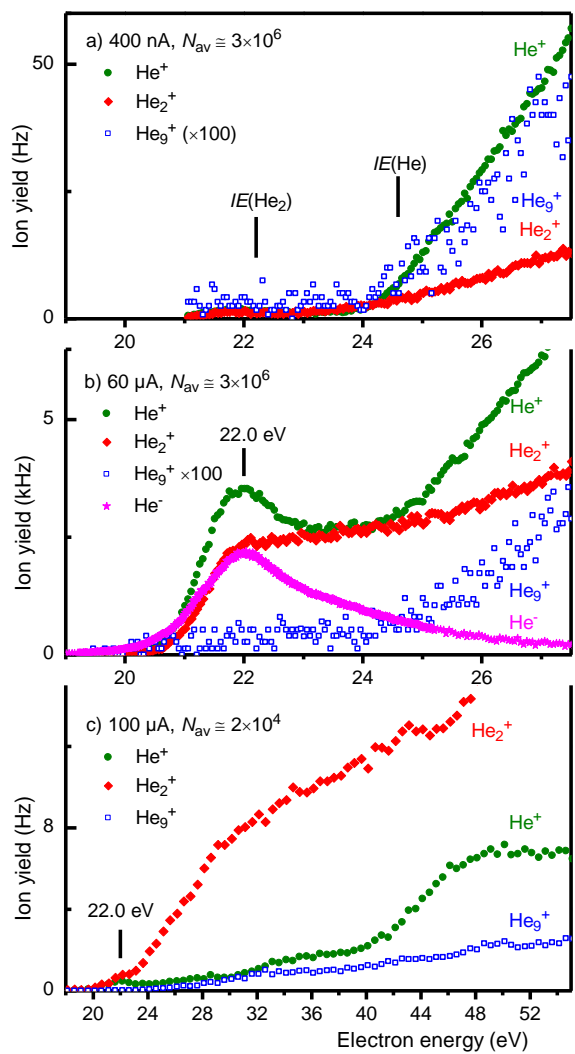

Fig. 1a: Ion yield of He$^+$, He$_2^+$ and He$_9^+$ versus electron energy recorded at low electron emission current (0.4 µA) for large helium droplets ($N_{av} \cong 3\times10^6$). Panel b: Same as panel a for large electron current (60 µA), together with the yield of He$^-$ anions measured under similar conditions. Panel c: Ion yields of He$^+$, He$_2^+$ and He$_9^+$ recorded at large electron current (100 µA) for small helium droplets ($N_{av} \cong 2\times10^4$) plotted on an extended energy scale. The adiabatic ionization energies of He and He$_2$ are marked by vertical lines in panel a; the feature at 22.0 eV is indicated in panels b and c.

Also shown in Fig. 1b is the yield of He$^-$ anions formed by electron attachment to droplets recorded without changing the source conditions or electron current.[36] An intense signal of He$^-$ is observed; its peak at 22.0 eV coincides with that of He$^+$ cations. He$_2^-$ (not shown) were also observed at a much smaller (factor 100) yield; the signal reached a maximum at 22.9 eV.[36]

Another dramatic change occurs when the helium droplet size is reduced by two orders of magnitude (Fig. 1c). A small maximum in the He$^+$ yield and a plateau in the He$_2^+$ yield near 22.0 eV are still discernible but the He$^+$ yield then remains essentially flat until the electron energy reaches 42 eV. This contrasts with the He$_2^+$ yield which increases strongly above 24 eV. Between 24 and 42 eV the He$^+$ yield is much less than that of He$_2^+$; it barely exceeds that of He$_9^+$.

In Figs. 2a and b the ion yields of He$^+$ and He$_2^+$ are plotted versus electron current $I_e$ for clusters of size $N_{av} = 3\times10^6$ and $1\times10^4$, respectively ($T_0 = 8.5$ and 10.5 K, respectively); the electron energy was 47 eV. Also shown is the yield of H$_2$O$^+$ which represents a minor impurity in the beam. The yields of He$_2^+$ and H$_2$O$^+$ are proportional to the electron current but He$^+$ varies with a higher power. This is demonstrated more clearly in Fig. 2c which displays the He$^+$ and He$_2^+$ data from Fig. 2b on a log-log scale. The solid lines represent the results of weighted least-squares fits of power laws to the ion yield $Y$,

$$Y(I_e) = aI_e^p + bI_e^q \qquad (2)$$

As observed above, the He$_2^+$ yield varies approximately linearly with electron current over the full range; the fitted exponent is $p = 1.14 \pm 0.01$ if parameter $b$ in eq. 2 is set to zero. The He$^+$ yield has a small linear component which is overwhelmed at high current by a non-linear component with power $q = 2.64 \pm 0.05$. A log-log plot of the data in Fig. 2a, and data recorded at an electron energy of 70 eV, exhibit similar trends.

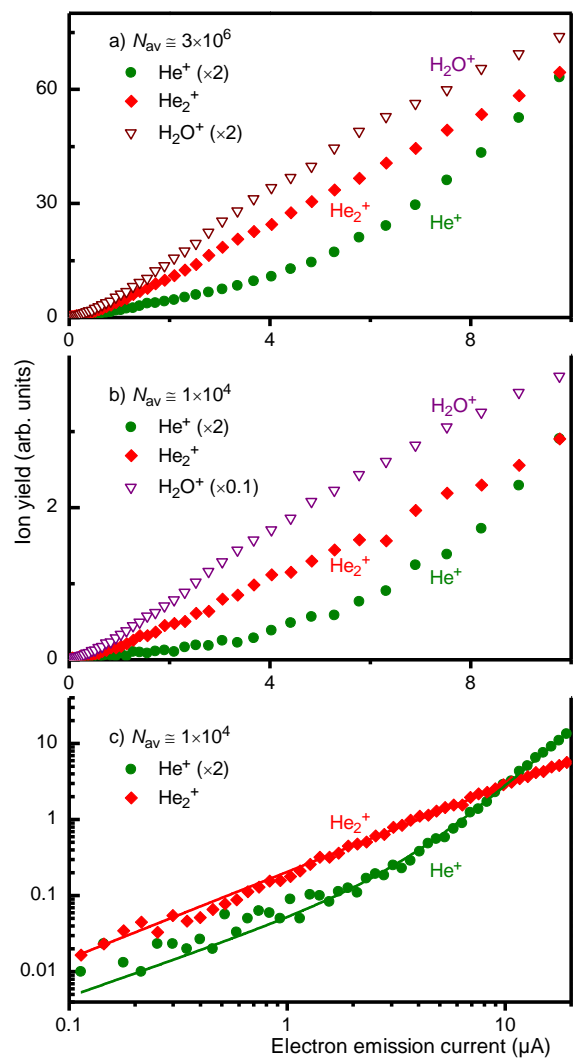

Fig. 2a: Ion yield of $He^+$, $He_2^+$ and $H_2O^+$ *versus* electron emission current recorded at an electron energy of 47 eV for large helium droplets ($N_{av} \cong 3\times10^6$). Panel b: Same as panel a for small droplets ($N_{av} \cong 1\times10^4$). Panel c: Data from panel b plotted on a double logarithmic scale; solid lines indicate weighted fits of a power law (eq. 2).

Before discussing these results we first review the interaction of electrons with helium. It will be important to distinguish between the formation of ions in the droplet and free ions. For example, the mass spectrometric detection of $He^+$ at subthreshold energies necessarily involves multiple excitation by two (or more) incident electrons but various pathways are conceivable, including: 1) formation of $He^+$ in the droplet by two incident electrons and subsequent ejection of $He^+$, 2) formation of metastable $He^*$ in the droplet by one electron, its subsequent ejection and ionization by another electron, or 3) excitation of a free helium atom in the background gas and subsequent ionization of $He^*$ by another electron. We will argue that only path (1) can account for the experimental observations.

The helium atom has a negative electron affinity but an electron may bind to electronically excited He. The only metastable helium anions $He^{*-}$ are those in quartet states; they may be viewed as an electron bound to $He^*$ in a triplet state (see [43] and references therein). The highest electron affinity, 77.5 meV, has been calculated for the lowest triplet state ($1s2s\ ^3S_1$); the longest lifetime of free $He^{*-}$ anions (359 μs) has been measured for the $1s2s2p\ ^4P_{5/2}$ state.[44]

Large $He_n^-$ cluster ions have been observed upon electron attachment to undoped helium droplets.[15-18] Resonance-like maxima appear in the anion signal at electron energies 2, 22, and 44 eV; the exact values depend on droplet size.[17] The resonance at 2 eV has been attributed to electrons that have barely enough energy to enter the droplet (the bottom of the conduction band of bulk helium lies 1.1 eV above the vacuum level), are localized and self-trapped in a bubble state. In bulk helium the energy of the bubble lies about 0.1 eV above the vacuum level; its radius is 17 Å.[23] The smallest helium droplet that supports a bubble state has been computed to contain 5200 helium atoms.[23] Experiments showed that the lifetime of bubble states decreases sharply as the droplet size falls below $10^6$ atoms;[18] the smallest observable cluster anions contained $7.5\times10^4$ helium atoms.[17]

The resonance at 22 eV in the yield of these large helium cluster anions has been attributed to an electron that excites an atom in the droplet to the lowest triplet state which, in vacuum, has an energy of 19.82 eV and a lifetime of 8000 s.[17] The inelastically scattered electron is then self-trapped in a bubble while He$^*$ forms a separate bubble with a radius of about 10 Å.[17] The energy difference between the first and second resonance (i.e. 22 - 2 = 20 eV) does, indeed, match the energy of free He$^*$ (19.82 eV). The resonance at 44 eV is assigned to sequential formation of two separate He$^*$ followed by self-trapping of the electron. Fine structure in the resonances at 22 and 44 eV has been attributed to He$^*$ excited into higher-lying states.[17]

Recently our group identified He$^-$ and He$_2^-$ anions following electron attachment to large helium droplets.[36] He$^-$ showed an onset at 21 eV and a maximum yield at $22.0 \pm 0.2$ eV followed by two unresolved resonances at 23.0 and 25.1 eV; a second group of resonances was observed at and above 44.0 eV with an onset around 42 eV. The energy dependence of He$^-$ was, except for the "missing" signal at 2 eV, similar to that of large He$_n^-$ cluster anions.[17] He$_2^-$ was observed at a two orders of magnitude smaller yield; its first maximum was located at 22.9 eV.

Mauracher et al.[36,37] attributed the resonant formation of He$^-$ at 22 eV to the following reaction: An incident electron enters the helium droplet and excites a helium atom into the $1s2s\ ^3S_1$ state. Within a few picoseconds the inelastically scattered electron thermalizes and becomes trapped in a bubble state. The highly polarisable He$^*$ and the electron bubble move towards each other and combine to form the heliophilic, highly mobile He$^{*-}$ in a quartet state.[43] The metastable ion is ejected if another incident electron is trapped in the droplet, thus explaining the approximately quadratic dependence of the He$^{*-}$ yield on the electron current.

We adopt and extend the model summarized above to arrive at the following consistent explanation for the observation of bare He$^+$ at subthreshold energies:
1. Formation of a highly mobile He$^{*-}$ by an incident electron as discussed above.[36,37]
2. Formation of He$^*$ by another incident electron.
3. Motion of He$^{*-}$ toward He$^*$ followed by Penning ionization (reaction 1). He$^*$ is heliophobic and resides on the surface; the large excess energy liberated in the Penning reaction suffices to eject He$^+$ from the droplet.

The probability for double excitation by two incident electrons increases, relative to the probability for single excitation, with increasing droplet size and electron current, thus explaining the strong subthreshold signal in Fig. 1b. For the time being we refrain from speculating about details of the He$^+$ ejection but the process may be akin to the well-studied formation of an intense He$_4^{*+}$ signal upon fusion of excimers in helium droplets.[12,31,45] Measurements of the kinetic energy release combined with molecular dynamics simulations, along the lines of a recent study of Ag$^*$ ejection from helium droplets,[46] would provide valuable insight.

The situation becomes more complex above 24.59 eV. Although a single incident electron may form He$^+$ in the droplet by direct ionization, this ion cannot escape from the droplet. Instead, after a few resonant hops the charge will self-trap to form He$_2^+$. Electrostriction combined with the high vibrational energy of the nascent He$_2^+$ will lead to the ejection of a small helium cluster ion containing as few as two or as many as hundreds of helium atoms.[12,22,24-27] The above-threshold signal of He$^+$ in Fig. 1a, recorded at low electron current, primarily originates from direct ionization of free He atoms in the background gas.

Another striking observation is the low He$^+$ yield in Fig. 1c for electron energies between 24 and 42 eV; the He$^+$ yield is hardly larger than that of He$_9^+$. Why? In this data set the average droplet size, $N_{asv} = 2\times10^4$, was four times smaller than the smallest droplet size that can support an electron bubble.[17] Thus, even though the electron current (100 µA) was very high, reaction (1) cannot occur.[†] Of course, the partial pressure of He atoms in the background gas in this run must have been small. Unfortunately this quantity is difficult to control. Apart from the droplet size the alignment of the molecular beam (i.e. nozzle, skimmer, and collimators) will play a role; the alignment may change with nozzle temperature.

Above 42 eV the He$^+$ yield shown in Fig. 1c exhibits a stepwise increase. At approximately this energy a single electron may form 2 He$^*$ + e$^-$ which, in principle, could lead to reaction (1). However, we had already ruled out formation of an electron bubble for droplets this small. At somewhat higher energies a single electron may form two $He^+$ in the droplet with subsequent ejection of $He^+$ due to Coulomb repulsion. The efficiency of this process is questionable though given the rapid conversion of $He^+$ to $He_2^+$ upon charge trapping. Furthermore, Fig. 2c indicates that the main contribution to the $He^+$ yield at 47 eV and high current is due to multiple excitations. Thus, although a single electron may produce two $He^*$, the formation of free $He^+$ involves two (or more) incident electrons. Another reaction mechanism, different from the ones discussed above for energies below 42 eV, seems to be at work in this regime.

Finally, we briefly address our results for $He_2^+$. Its yield increases abruptly above the thermodynamic threshold even for small droplets (Fig. 1c) and varies linearly with electron current (Fig. 2c). These results are in accord with the accepted model for formation of small $He_n^+$ upon electron ionization, namely formation of $He^+$ by direct ionization in the droplet, and resonant charge hopping that ends in the formation of a highly excited $He_2^+$ which is ejected.[12,22,24-27]

**Conclusion**

We have shown that free $He^+$ is produced by electron ionization of helium droplets at subthreshold energies, and we have proposed a detailed formation mechanism. An incident electron will form $He^*$ plus an electron bubble; the pair will react to form $He^{*-}$. If another $He^*$ is present the highly mobile $He^{*-}$ will move toward it and produce $He^+$ which is ejected as a result of the highly exothermic reaction in the surface region.

The validity of the model may be tested by photoionization at subthreshold energies because this way the required intermediate, an electron that is self-trapped in a bubble and subsequently reacts with $He^*$ to form $He^{*-}$, is absent. In fact, no significant $He^+$ signal was observed at subthreshold energies in a recent photoionization study by Buchta et al.[14] even though small $He_n^+$ did form as a result of multiple excitations. However, the droplets in that study contained fewer than $10^4$ helium atoms; the proposed model is not applicable to droplets that small.

**Acknowledgements**

This work was supported by the Austrian Science Fund, Wien (FWF Projects I978, P23657, and P26635).

**Footnotes**

† Strictly speaking reaction (1) would be strongly suppressed for a mean size of $2 \times 10^4$ but not impossible. The size distribution of the helium droplets follows a log-normal distribution; a fraction of the droplets will be much larger than the mean.